# Nebula-Relay Hypothesis: The Chirality of Biological Molecules in Molecular Clouds

Lei Feng[1,2,3]*

[1]Key Laboratory of Dark Matter and Space Astronomy, Purple Mountain Observatory, Chinese Academy of Sciences, Nanjing, China, [2]School of Astronomy and Space Science, University of Science and Technology of China, Hefei, China, [3]Joint Center for Particle, Nuclear Physics and Cosmology, Nanjing University - Purple Mountain Observatory, Nanjing, China

The chiral puzzle of biological molecules is thought to be closely related to the origin of life and is still a mystery so far. Previously, we proposed a new model on the origin of life, Nebula-Relay hypothesis, which assumed that the life on Earth originated at the planetary system of the Sun's predecessor star and then filled in the pre-solar nebula after its death. As primitive lives existed in the pre-solar nebula for a long time, did the chiral biomolecules form during this period? We explore such a possibility in this work and find that the ultra-low temperature environment of molecular clouds is beneficial to generating the chiral polymer chain of biological molecules.

## 1 INTRODUCTION

Symmetry usually refers to the invariance of an object under some transformations, such as translation, reflection, and rotation. It is also a fundamental aspect of biological molecules. In chemistry, if a molecule or ion cannot be superposed on its mirror image by any transform, we call them enantiomers. These two enantiomers are usually labeled as "right(left)-handed" or D(L) forms. It is the nature of chirality and an intrinsic property of natural chemical molecules. Unlike typical chemical synthesis, chirality is broken in the biochemical processes of living systems. For example, there are only L-amino acids in proteins and D-ribose in ribonucleic acids for almost all creatures on Earth. The origin of biological chiral is an intractable mystery that has plagued us for a long time and is thought to be related to the origination of life.

Generally speaking, there are two most persuasive and popular models, i.e., abiogenesis (Alexander Oparin, 2021; Haldane, 2021) and panspermia theory (Davies, 1988). The famous Miller–Urey experiment (Miller, 1953) proved that amino acids (constituents of the proteins) could be synthesized in the reaction chamber, which simulates the environment of the early earth. However, equivalent left- and right-handed amino acids are produced in this experiment. Some new experiments demonstrated the catalytic action of chondritic meteorite minerals in mild aqueous environments (Rotelli et al., 2016; Cabedo et al., 2021), which is very instructive for this field. Alexander et al. (2017) and Alexander et al. (2018) pointed out that the insoluble organic materials in comets are potential sources of volatiles for the terrestrial planets in the solar system. Anyway, the origin of life is much more complicated than we thought, and new mechanisms need to be introduced into the origination model of life.

There are many discussions about chiral puzzle already. One possible explanation is proposed by Vester–Ulbricht, who assume that left-handed electrons, produced by parity-violating beta decays, interact with biological precursor molecules and destroy one type of enantiomer more significantly (Vester et al., 1959; Ulbricht and Vester, 1962) than another. The circularly polarized bremsstrahlung produced by the longitudinally spin-polarized beta electrons, they guess, is the reason which preferentially photolyzes certain chiral molecules (Ulbricht and Vester, 1962).

Some studies pointed out that the interaction between beta electrons and chiral molecules can induce a left–right asymmetry of order $\sim 10^{-6}$ (Zeldovithe and Sakyan, 1980; Wang and Luo, 1985). Several experiments were designed to search evidence of the above scenario, but no one achieved positive and repeatable results (Bonner et al., 1979; Hodge et al., 1979; Bonner, 1991; Bonner, 2000; Fitz et al., 2007). Circularly polarized light has been detected in molecular clouds. For example, the authors reported a high degree of circular polarization of as much as 22% in NGC 6334 in Kwon et al. (2013). According to the Vester–Ulbricht hypothesis, chiral amino acids can be synthesized under such an environment.

Carbonaceous chondrites are the only rocks that can provide original records of pre-solar nebula processes. The famous Murchison meteorite is one kind of carbonaceous chondrite which is rich in organic compounds and contends more than fifteen types of amino acids (Wolman et al., 1972). Some types of amino acids were found to have an excess of the L-configuration (Engel and Nagy, 1982; Cronin and Pizzarello, 1997; Engel and Macko, 1997), but there are also many objections (Bada et al., 1983; Pizzarello and Cronin, 1998). Schmitt-Kopplin and his collaborators had identified 70 amino acids in Murchison meteorite (Schmitt-Kopplin et al., 2010). Several amino acids are also distinguished from a carbonaceous chondrite at Mukundpura, India (Rudraswami et al., 2019). There are ninety two types of amino acids identified in meteorites according to current research studies (Cronin et al., 1995; Martins et al., 2009; Glavin et al., 2010; Elsila et al., 2016; Burton and Berger, 2018). These facts may hint that there are amino acids in molecular clouds.

This paper is organized as follows: In **Section 2**, we briefly introduce the Nebula-Relay hypothesis. Then, we present the calculation method we used in this manuscript in **Section 3**. The main results are presented in **Section 4**, and the conclusions are summarized in **Section 5**.

## 2 NEBULA-RELAY HYPOTHESIS

In panspermia theory, it is presumed that primitive lives spread to Earth from other planets in the solar system or other planetary systems in the Milky Way. However, it does not solve the problem of how and where life originated, let alone the chiral puzzle of biological molecules.

Feng (2021) proposed a new hypothesis on the origin of Earth's life, named "Nebula-Relay hypothesis" or "local panspermia." The original intention of our hypothesis is that life on Earth may appear too early (Dodd et al., 2017). We assume that the chemical origin theory is correct but takes billions of years. Then, we believe that primitive life formed on the planet system of the sun's predecessor star and filled in the produced pre-solar nebula after the death of the predecessor star. So, the history of Earth's life can be divided into three epochs: the origin of life on the planetary system of the Sun's predecessor star, primitive life epoch in the pre-solar nebula, and Earth life epoch.

In epoch I, primitive creatures were formed on the planet of the pre-solar star through complex physicochemical interactions. The environment of the pre-solar planet may be similar to that of the early Earth, and then explorations on the origin of Earth's life are inspiring and informative. Rotelli et al. (2016) pointed out that most of the complex organic compounds were produced by the catalytic action of minerals under the presence of heat and water. Complex organics in meteorites can also be generated in their parent asteroids through Fischer–Tropsch catalytic reactions (Cabedo et al., 2021). Nuth et al. (2020) studied the complex carbon cycle on the evolving asteroid through thermal or hydrous metamorphic processes.

Cosmic ray (CR, also including the contribution of nearby bright stars) particles collide with the gas in nebula and produce low-energy secondary particles. It reduces high-energy CR particles and protects primitive life from lethal radiation. So, we propose that low-energy secondary electrons and (or) photons of CR are the energy source of living creatures in molecular clouds (Feng, 2021).

There are differences between this model and ordinary panspermia theory. Different from the shuttle of seeds in panspermia, primitive life in this model occurred on the local place at the pre-solar epoch and scattered in the pre-solar nebula. Homologous lives are almost everywhere in the solar system, and we can search for the traces of these homologous creatures to test this model. As the energy comes from the cosmic ray and the encapsulation of the meteorite prevents this process, there are no living forms in meteorite materials. Although challenging, their fossils might be found in meteorites.

Can the chiral puzzle be satisfactorily explained in the Nebula-Relay hypothesis? We explore this problem in this article and try to prove that the low-temperature environment is conducive to the production of biological chirality.

## 3 SOME PHYSICAL BACKGROUNDS

The previous section shows that the chiral biomolecules could be achieved through their interaction with longitudinally spin-polarized beta electrons or circularly polarized light. However, no significant chiral differences were found in the amino acids of meteorites. In other words, the difference between enantiomers is tiny.

This work aims to achieve the chirality of biomolecules with such a slight left–right deviation. Following the calculation method in Luo (1994), we try to explore the amplification of left–right asymmetry through the polymerization process in the low-temperature environment of molecular cloud, i.e., the production of homochiral polymers.

Firstly, we suppose that the distribution of chiral monomers satisfies the Boltzmann distribution, i.e., $n_{L,D} \sim \exp(-E_{L,D}/K_B T)$, where $E_{L,D}$ is the energy of L(D)-form chiral monomer, respectively, and $K_B$ is the Boltzmann constant. The following formula can describe the polarization of chiral monomers:

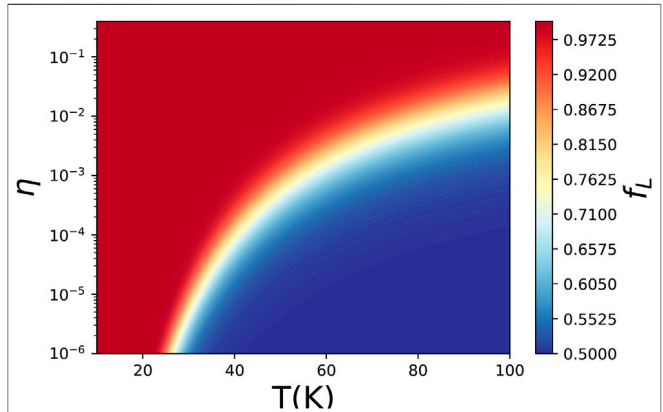

FIGURE 1 | Probability of L-type monomer in the polymer chain.

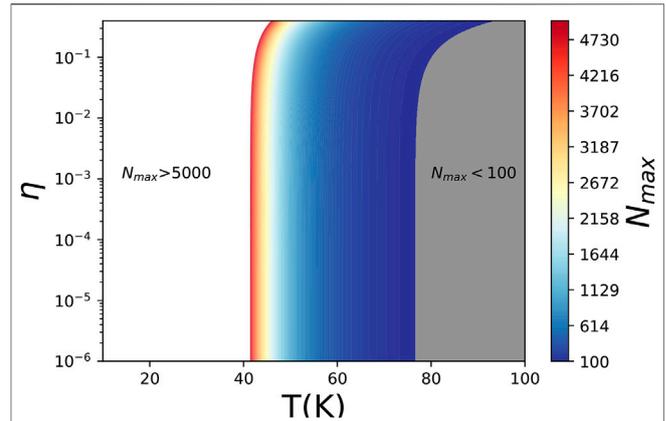

FIGURE 2 | Dependence of the maximum pure L-type chain length N on $\eta$ and temperature.

$$\eta = (n_L - n_D)/(n_L + n_D) = \tanh J, \quad (1)$$

where

$$J = (E_D - E_L)/2K_BT. \quad (2)$$

Then, we focus on the recombined energy of neighboring monomers $E_{LL,DD,LD,DL}$ and set $E_{LL} = E_{DD}$, $E_{LD} = E_{DL}$ for symmetry consideration. The chiral interaction energy is defined as follows:

$$U = (E_{LD} - E_{LL})/2K_BT. \quad (3)$$

If $U > 0$, the system tends to state the identical chiral monomers stick together. We then define stereoselectivity $\gamma$ to describe the difference of probabilities of adding distinct chiral monomers as in Luo (1994), which is

$$\gamma = \tanh U = \frac{\exp(-E_{LL}/K_BT) - \exp(-E_{LD}/K_BT)}{\exp(-E_{LL}/K_BT) + \exp(-E_{LD}/K_BT)}. \quad (4)$$

To form a pure L-form polymer chain with length $N_c$ and relatively high probability, there must be (Avetisov et al., 1991)

$$\frac{(1-\eta)(1-\gamma)}{2(1+\eta\gamma)} < N_c^{-1}. \quad (5)$$

The Hamiltonian of the system is described by

$$H = -K_BT\left(J\sum_i \sigma_i + U\sum_i \sigma_i\sigma_{i+1}\right), \quad (6)$$

where the chirality of each monomer is described by the spin $\sigma_i$ $\sigma_i = \pm 1$ denotes L- and D-form chiral molecular). As in Luo (1994), a statistical ensemble with the above Hamiltonian can be used to calculate the equilibrium distribution of chiral monomers through the Ising model (Huang, 1963). The chiral polarization of polymer $f_L$ can be calculated through the following formula:

$$f_L = \frac{1}{2}\left\{1 + \sinh J/\left[\sinh^2 J + e^{-4U}\right]^{1/2}\right\}, \quad (7)$$

where $f_{L(D)}$ denotes the probability of L(D)-form monomer in the chain and $f_L + f_D = 1$. $f_L = 1$ denotes the pure L-form monomer and vice versa. It is obvious that if $\exp^{-4U} \ll \sh J$, $f_L \sim 1$.

Combining amino acids into peptide chains is an exothermic reaction in nature, and reactions between homochiral molecules release more heat. Therefore, more energy is required to decompose the bonds of homochiral molecules during the photolysis/radiolysis process, but L-D chemical bond is much easier to break down. This effect would further amplify pure L-form polymer chains, although we do not consider this effect here.

## 4 CHIRAL OF BIOMOLECULES IN A LOW-TEMPERATURE ENVIRONMENT

The typical temperature of molecular cloud is about 10 ~ 50 K but has significant differences in different regions. Wilson et al. (1997) pointed out that molecular clouds with H II regions have temperatures of 15–100 K, and even larger in the giant H II region. Therefore, we set 10 K < T < 100 K and $E_{LD} - E_{LL}$ = 700 Cal/mol (1% of C-N bound energy) in following calculations.

The probability of L-form monomer in the polymer chain is shown in **Figure 1**. It is easy to see that low temperature and large polarization of chiral monomers could enlarge the probability of an L-form monomer. At low temperature ($T < 25$ K), a tiny deviation between L and D types of monomers could induce a left-handed chain, which is consistent with the critical temperature in Luo (1994) with polarization $10^{-6}$. For $\eta \sim 10^{-2}$, temperatures below 100 K are all suitable.

According to **Eq. 5**, we define the maximum permissible length of pure L-type polymer chain $N_{max}$ as follows:

$$N_{max} = \left[\frac{2(1+\eta\gamma)}{(1-\eta)(1-\gamma)}\right]. \quad (8)$$

The relation between $N_{max}$ and $(\eta, \gamma)$ is shown in **Figure 2**. From the figure, we can see that the maximum length of pure

L-type chain is less than 100 as $T > 80$ K because $\eta$ and $\gamma$ decrease sharply with increasing temperature.

During the formation of the solar system, the temperature increased in high-density areas. But the temperature in the other regions is still relatively low. Moreover, the left-handed biological system has already formed, and temperature changes no longer affect the chirality of biomolecules.

## 5 SUMMARY AND DISCUSSIONS

In this article, we calculate the probability of generating homochiral polymers in the nebula environment with a slight deviation in the density of the left- and right-handed monomers. We intend to explain the chiral puzzle of biomolecules. Our calculations show that chiral polymers can be naturally induced in an ultra-low temperature molecular cloud environment because the polarization and stereoselectivity of chiral biomolecules are much smaller than the homeothermy one. Moreover, a pure left-handed polymer chain can be much longer in a low-temperature environment. Furthermore, this model might be tested through low-temperature experiments.

## DATA AVAILABILITY STATEMENT

The original contributions presented in the study are included in the article/Supplementary Materials, further inquiries can be directed to the corresponding author.

## AUTHOR CONTRIBUTIONS

The author confirms being the sole contributor of this work and has approved it for publication.

## FUNDING

This work was supported by the National Natural Science Foundation of China (Grant No. 11 773 075) and the Youth Innovation Promotion Association of Chinese Academy of Sciences (Grant No. 2016 288).

## ACKNOWLEDGMENTS

The author thanks Dr. Lei Zu and Zi-Qing Xia for their generous help in perfecting this article.